\documentstyle[12pt]{article}

\def\beq{\begin{equation}}
\def\eeq{\end{equation}}
\def\bea{\begin{eqnarray}}
\def\eea{\end{eqnarray}}

\def\ba{\begin{array}}
\def\ea{\end{array}}
\def\ds{\displaystyle}

\def\,{\"{U}}
\def\6{\.{I}}

\begin{document}

\title{Supersymmetric Solutions of PT-/non-PT-Symmetric and non-Hermitian
central Potentials via Hamiltonian Hierarchy Method}
\author{Gholamreza Faridfathi, Ramazan Sever\thanks{%
Corresponding author: sever@metu.edu.tr}, Metin Akta\c{s} \\
Department of Physics, Middle East Technical University, 06531 Ankara, Turkey}
\date{\today}
\maketitle

\begin{abstract}
The supersymmetric solutions of PT-/non-PT-symmetric and
non-Hermitian deformed Morse and P\"{o}schl-Teller potentials are
obtained by solving the Schr\"{o}dinger equation. The Hamiltonian
hierarchy method is used to get the real energy
eigenvalues and corresponding eigenfunctions.
\end{abstract}

\baselineskip0.9cm\bigskip


\newpage

\section{Introduction}

PT-symmetric quantum systems have generated much interest in
recent years [1]. About ten years ago, Besis suggested that the
eigenvalue spectrum of complex-valued Hamiltonians is real and
positive. Bender and Boettcher claimed that this result is due to
PT-symmetry where P and T are the parity and time reversal
operators respectively. It is neither a necessary nor a sufficient
condition for a Hamiltonian to have real spectrum. In particular,
the spectrum of the Hamiltonian is real if PT-symmetry is not
spontaneously broken. Thus, the property of exactness guarantees
the real eigenvalues. Recently, Mostafazadeh introduced another
concept for a class of PT-invariant Hamiltonians called $\eta
-pseudo-Hermiticity$ [2]. In fact, Hamiltonians of this type
satisfy the transformation $\eta ~\hat{H}~\eta ^{-1}=\hat{H}^{\dag
}$ [3]. Moreover, completeness and orthonormality conditions for
the eigenstates of such potentials are proposed [4]. Various
techniques have been applied in the study of PT-invariant
potentials such as variational methods [5], numerical approaches
[6], Fourier analysis [7], semi-classical estimates [8], quantum
field theory [9] and Lie group theoretical approaches [10-13]. In
the applications, a generalization of the symmetry concept is
encountered in the supersymmetric quantum mechanics (SUSYQM) [14].
A variety of PT-symmetric examples can be found using the SUSYQM
techniques [15-20]. Furthermore, one can get more examples of the
PT-symmetric and non-PT-symmetric and also non-Hermitian potential
cases such as oscillator type potentials [21], flat, step and
double square-well like potentials within the framework of SUSYQM
[22, 23], exponential type screened potentials [24],
quasi/conditionally exactly solvable ones [25], complex
H\'{e}non-Heiles potentials [26], periodic isospectral potentials
[27] and some others[28, 29].

The aim of the present work is to calculate the energy eigenvalues
and the corresponding eigenfunctions of the deformed Morse and P%
\"{o}schl-Teller potentials using the Hamiltonian hierarchy method
[30] within the framework of the PT-SUSYQM. This method is also
known as the factorization method introduced by Schr\"{o}dinger
[31] and later developed by Infeld and Hull [32]. It is useful to
obtain the energy spectra for different potentials in
non-relativistic quantum mechanics [33, 34, 35].

The organization of this paper is as follows: In Sec. II we
introduce a brief review of the Hamiltonian hierarchy method. In
Secs. III and IV we present the supersymmetric solutions of
PT-symmetric and Hermitian/non-Hermitian forms of the well-known
potentials by using this method. We discuss the results in Sec. V.

\section{Hamiltonian Hierarchy Method}

The radial Schr\"{o}dinger equation for some specific potential
energies can be solved analytically only for the states with zero
angular momentum [36, 37]. However, in supersymmetric quantum
mechanics, one can get exact results with the hierarchy problem by
using effective potentials for non-zero angular momentum states.
In the framework of SUSYQM, this method provides an eigenvalue
spectra for adjacent members of supersymmetric partner
Hamiltonians. These Hamiltonians share the same eigenvalue spectra
except for the missing ground state.

In the application of this method, we first look for an effective
potential similar to the original specific potential and inspired
by the SUSYQM to propose a superpotential, namely $W_{\left(
l+1\right) }(x)$, as an ansatz, where $\left( l+1\right) $ denotes
the partner number with $l=0,1,2...$. Substituting the proposed
superpotential into the Riccati equation,

\begin{equation}
V_{\left( l+1\right) }(x)-E_{(l+1)}^{0}=W_{\left( l+1\right)
}^{2}(x)-\frac{dW_{(l+1)}(x)}{dx},
\end{equation}

\noindent the $\left( l+1\right) $th member of the Hamiltonian
hierarchy can be obtained. Taking into account the shape
invariance requirement [14], the bound-state energies can be
obtained through Eq. (1), and the corresponding eigenfunctions by
means of,

\begin{equation}
\Psi_{(l+1)}(x)=N\exp (-\int^{r}W_{(l+1)}(x^{\prime })dx^{\prime }).
\label{e9}
\end{equation}

\section{Generalized Morse Potential}

The generalized Morse potential is given by [24]

\bigskip

\begin{equation}
V(x)=V_{1}e^{-2\alpha x}-V_{2}e^{-\alpha x}.  \label{e1}
\end{equation}
To apply the Hamiltonian hierarchy method, we shall take the coefficients
as $V_{1}=V,$ and $\ds{\frac{V_{2}}{V_{1}}=q}$. Inspired by
the SUSYQM, we propose an ansatz for the superpotential



\begin{equation}
W_{\left( l+1\right) }(x)=-\lambda e^{-\alpha x}+(\lambda q-\frac{2l+1}{2}),
\label{e4}
\end{equation}

\bigskip

\noindent where, $\ds{\lambda ^{2}=\frac{2mV}{a^{2}\hbar ^{2}}}$, $\ $and
$(2l+1)$
denotes the partner number with $l=0,1,2,...,$ and the parameter $m$ is the
reduced mass of a diatomic molecule. The superpotential chosen in Eq.(4)
leads to the $(l+1)th$ member of the Hamiltonian hierarchy  through
the Riccati equation as,

\begin{equation}
V_{\left( l+1\right) }(x)-E_{(l+1)}^{0}=W_{\left( l+1\right) }^{2}(x)-\frac{1%
}{\alpha }\frac{dW_{(l+1)}(x)}{dx},  \label{e5}
\end{equation}
which yields,

\begin{equation}
V_{(l+1)}(x)=\lambda ^{2}(e^{-2\alpha x}-qe^{-\alpha x})+2l\lambda
e^{-\alpha x}.  \label{e6}
\end{equation}
Now, using the shape invariance requirement, the energy
eigenvalues for any n-th state become,


\begin{equation}\label{e8}
E_{(l+1)}^{n}=-(\lambda q-\frac{2l+n+1}{2})^{2},\quad \quad
n=0,1,2,\ldots
\end{equation}
The corresponding eigenfunctions are obtained through Eq. (2) as

\begin{equation}
\Psi _{\left( l+1\right) }^{n=0}(x) =N\exp \left[ -\frac{\lambda }{\alpha }%
e^{-\alpha x}-(\lambda q-\frac{2l+1}{2})x\right] ,  \label{e10}
\end{equation}
where, N is the normalization constant.

\bigskip

\subsection{Non-PT symmetric and non-Hermitian Morse Case}

\bigskip

We define the potential parameters in Eq. (3) as $V_{1}=\left(
A+iB\right) ^{2}$, \ \ $ V_{2}=\left( 2C+1\right) \left(
A+iB\right) ,$ and $\alpha =1$. $A,$ $B$ and $C$ are real, and $\
i=\sqrt{-1}.$ For simplicity we define $A+iB=i\omega $, $\left(
A+iB\right) ^{2}=-\omega ^{2}$, $2C+1=K$. Thus, the potential
becomes

\bigskip

\begin{equation}
V(x)=-\frac{\omega ^{2}}{K}\left[ Ke^{-2x}-\frac{K^{2}}{i\omega }e^{-x}%
\right] .  \label{e12}
\end{equation}
To get the final compact form, we also define $\ds{\frac{\omega
^{2}}{K}=G}$, and $\ds{\frac{K^{2}}{\omega }=t}$, and $GK=D$, and
also $\ds{\frac{t}{K}=P}$. As a result, we get,

\begin{equation}
V\left( x\right) =-D\left[ e^{-2x}+iPe^{-x}\right] .  \label{e13}
\end{equation}
We propose an ansatz for the superpotential as,

\begin{equation}
W_{\left( l+1\right) }(x)=-i~\lambda e^{-x}+(\lambda -\frac{2l+1}{2}),
\label{e15}
\end{equation}
\noindent where $\ds{\lambda ^{2}=\frac{2mD}{a^{2}\hbar ^{2}}}$.
Consequently, according to the Hamiltonian hierarchy method, we
get,

\begin{equation}
V_{\left( l+1\right) }(x)=-\lambda ^{2}(e^{-2x}+2i~e^{-x})+2il\lambda
e^{-x},
\label{e16}
\end{equation}
By substituting this into Eq. (1), the corresponding eigenvalues
and eigenfunctions are obtained as

\begin{equation}
E_{\left( l+1\right) }^{n}=-(\lambda -\frac{n+2l+1}{2})^{2},\quad \quad
n=0,1,2,\ldots ,  \label{e17}
\end{equation}
and
\begin{equation}
\Psi _{\left( l+1\right) }^{n=0}(x)=N\exp \left[ -i\lambda e^{-x}-(\lambda
-\frac{2l+1}{2})x\right] ,  \label{e18}
\end{equation}
where, N is a normalization constant. \bigskip

\subsection{The first type of PT-symmetric and non-Hermitian Morse case}

\bigskip
We take the coefficients of the generalized Morse potential as
$V_{1}=\left( A+iB\right) ^{2}$, \ \ $V_{2}=(2C+1)(A+iB)$, and
$\alpha=i$. Following the same procedure, we propose an ansatz for
the superpotential


\begin{equation}
W_{\left( l+1\right) }(x)=-\lambda e^{-ix}+(\lambda
-\frac{2l+1}{2}). \label{e20}
\end{equation}
\noindent The Hamiltonian hierarchy method yields,

\begin{equation}
V_{\left( l+1\right) }(x)=\lambda ^{2}(e^{-2ix}-e^{-ix})+2l\lambda e^{-ix}.
\label{e21}
\end{equation}

\noindent This form of potential gives the same eigenvalues as in
Eq. (13).
\bigskip


\subsection{The second type of PT-symmetric and non-Hermitian Morse case}

\bigskip

Now, we take the parameters as $V_{1}=-\omega ^{2},$ and
$V_{2}=D$, and $\alpha =i\alpha$ in Eq. (3), where $V_{1}$ and
$V_{2}$ are real. For $V_{1}\Longrightarrow 0 $, we get no real
spectra for this kind of PT-symmetric Morse potentials. The
superpotential can be proposed as,

\bigskip
\begin{equation}
W_{\left( l+1\right) }(x)=-e^{-i\alpha x}+(2l+1+\frac{D}{2\omega }).
\label{e24}
\end{equation}
\bigskip By applying the Hamiltonian hierarchy method, we get the
potential

\begin{equation}
V_{(l+1)}(x)=e^{-2i\alpha x}-2\left[ \left( 2l+1\right) +\frac{D}{2\omega }+%
\frac{i\alpha }{2}\right] e^{-i\alpha x},  \label{e25}
\end{equation}
and the corresponding eigenvalues for any n-th state are,


\begin{equation}
E_{(l+1)}^{n}={-(2l+n+1+\frac{D}{2\omega })}^{2}.  \label{e27}
\end{equation}

\bigskip

\section{P\"{o}schl-Teller Potential}

\bigskip

The P\"{o}schl-Teller potential is given as,

\begin{equation}
V\left( x\right) =-4V_{0}\frac{e^{-2\alpha x}}{\left( 1+qe^{-2\alpha
x}\right) ^{2}}.  \label{e28}
\end{equation}
In the framework of the SUSYQM, the corresponding superpotential
can be proposed as,

\begin{equation}
W_{\left( l+1\right) }\left( x\right) =-\frac{\hbar }{\sqrt{2m}}\frac{\left(
l+1\right) e^{-2\alpha x}}{\left( 1+qe^{-2\alpha x}\right) ^{2}}+\sqrt{\frac{%
m}{2}}\frac{e^{2}}{\hbar }\left[ \frac{1}{\left( l+1\right) }-\frac{\left(
l+1\right) }{2}\beta \right] ,  \label{e29}
\end{equation}
where, ${\displaystyle\beta =\frac{\hbar ^{2}}{me^{2}},}$ and $l=0,1,2,...$%
\noindent . By applying the Hamiltonian hierarchy method we get,

\begin{equation}
V_{\left( l+1\right) }(x)=\frac{\hbar ^{2}}{2m}\frac{e^{-4\alpha x}}{\left(
1+qe^{-2\alpha x}\right) ^{4}}l(l+1)-e^{2}\frac{e^{-2\alpha x}}{\left(
1+qe^{-2\alpha x}\right) ^{2}}\left[ 1-l\left( l+1\right) \frac{\beta }{2}%
\right] .  \label{e30}
\end{equation}
As a result, the corresponding eigenvalues of this potential for any n-th
state are,


\begin{equation}
E_{\left( l+1\right) }^{n}=-\frac{q^{2}me^{4}}{2\hbar ^{2}}\left[ \frac{1}{%
\left( n+l+1\right) }-\frac{\left( n+l+1\right) }{2}\beta
\right]^{2} . \label{e32}
\end{equation}
The corresponding eigenfunctions are,

\begin{equation}
\Psi _{\left( l+1\right) }^{n=0}(x)=N\left( 1+qe^{-2\alpha x}\right)
^{l+1}\exp \left\{ -\frac{me^{2}}{\hbar ^{2}}\left[ \frac{1}{l+1}-\frac{%
\left( l+1\right) }{2}\beta \right] x\right\} ,  \label{e33}
\end{equation}
where, N is a normalization constant. \bigskip

\subsection{Non-PT symmetric and non-Hermitian P\"{o}schl-Teller cases}

Here, $V_{0}$ and $q$ are complex parameters:
$V_{0}=V_{0R}+iV_{0I}$
\ and \ $q=q_{R}+iq_{I}$ , but $%
\alpha $ is a real parameter. Although the potential is complex
and the corresponding Hamiltonian is non-Hermitian and also non-PT
symmetric, there may be real spectra if and only if $\ \ V_{0I}$ $%
q_{R}=V_{0R}$ $q_{I}.$ When both parameters $V_{0},$ and $q$ are taken pure
imaginary, the potential turns out to be,\

\begin{equation}
V\left( x\right) =-4V_{0}\frac{2qe^{-4\alpha x}+i(1-q^{2}e^{-4\alpha x})}{%
\left( 1+q^{2}e^{-4\alpha x}\right) ^{2}}.  \label{e34}
\end{equation}
For simplicity, we use the notation $V_{0}$ and $q$ instead of
$V_{0I}$ and $q_{I}.$ To obtain the energy eigenvalues, we propose
the superpotential

\begin{equation}
W_{\left( l+1\right) }\left( x\right) =-\frac{\hbar }{\sqrt{2m}}\frac{\left(
l+1\right) qe^{-4\alpha x}}{\left( 1+q^{2}e^{-4\alpha x}\right) ^{2}}+\sqrt{%
\frac{m}{2}}\frac{e^{2}}{\hbar }\left[ \frac{1}{\left( l+1\right) }-\frac{%
\left( l+1\right) }{2}\beta \right] .  \label{e35}
\end{equation}
Therefore, substituting this equation into the Riccati equation,
we get the same potential as in Eq. (22), and also the same energy
eigenvalues as in Eq. (23).

\bigskip

\subsection{PT symmetric and non-Hermitian P\"{o}schl-Teller case}

\bigskip

We choose the parameters $V_{0}$ and $q$ as real, and also $\alpha
\Longrightarrow i\alpha$ in Eq. (25). Here, we propose the
superpotential similar to this potential as,

\begin{equation}
W_{\left( l+1\right) }\left( x\right) =-\frac{\hbar }{\sqrt{2m}}\frac{\left(
l+1\right) qe^{-4i\alpha x}}{\left( 1+q^{2}e^{-4i\alpha x}\right) ^{2}}+%
\sqrt{\frac{m}{2}}\frac{e^{2}}{\hbar }\left[ \frac{1}{\left( l+1\right) }-%
\frac{\left( l+1\right) }{2}\beta \right] .  \label{e37}
\end{equation}
By applying the same procedure, we get the energy eigenvalues as
in Eq.(23).

\section{Conclusions}

We have applied the PT-symmetric formulation to solve the
Schr\"{o}dinger equation for more general Morse and
P\"{o}schl-Teller potentials. The Hamiltonian hierarchy method
within the framework of the SUSYQM is used. We have obtained the
energy eigenvalues and the corresponding eigenfunctions for
different forms of these potentials. The energy spectrum of the
PT-invariant complex-valued non-Hermitian potentials may be real
or complex depending on the parameter values. We have imposed some
restrictions on the potential parameters to get the real spectra
in PT-symmetric, or more generally, in non-Hermitian cases. It is
also pointed out that the superpotentials, and partners must
satisfy the PT-symmetry condition. Finally, we have pointed out
that our exact results of complexified general Morse and
P\"{o}schl-Teller potentials may increase the number of
applications in the study of different quantum systems.


\newpage

\begin{thebibliography}{99}
\bibitem{ref1}  C. M. Bender, S. Boettcher, Phys. Rev. Lett. {\bf 80}%
, 5243(1998),\newline C. M. Bender, J. Math. Phys. {\bf 40},
2201(1999),\newline C. M. Bender, D. C. Brody, H. F. Jones, Phys.
Rev. Lett. {\bf 89}, 270401(2002),\newline
C. M. Bender, P. N. Meisinger, Q. Wang, J. Phys. A{\bf 36},6791(2003)%
\newline
C. M. Bender, D. C. Brody, H. F. Jones, Am. J. Phys. {\bf 71}, 1095(2003),%
\newline
C. M. Bender, P. N. Meisinger, Q. Wang, J. Phys A{\bf 36}, 1029(2003),%
\newline
C. M. Bender, M. V. Berry, A. Mandilara, J. Phys. A{\bf 35},L467(2002)

\bibitem{ref2}  A. Mostafazadeh, J. Math. Phys. {\bf 43}, 205(2002), ibid
{\bf 43}, 2814(2002), ibid {\bf 43}, 3944(2002), J. Phys. A{\bf
36}, 7081(2003)

\bibitem{ref3}  Z. Ahmed, Phys. Lett. A{\bf 290}, 19(2001), ibid {\bf 310},
139(2003), J. Phys. A{\bf 36}, 10325(2003)

\bibitem{ref4}  S. Weigert, Phys. Rev. A{\bf 68}, 062111(2003)

\bibitem{ref5}  C. M. Bender, F. Cooper, P. N. Meisinger, M. V. Savage,
Phys. Lett. A{\bf 259}, 224(1999)

\bibitem{ref6}  C. M. Bender, G. V. Dunne, P. N. Meisinger, Phys. Lett. A{\bf 252},
 272(1999)

\bibitem{ref7}  V. Buslaev, V. Grecchi, J. Phys. A{\bf 36}, 5541(1993)

\bibitem{ref8}  E. Delabaere, F. Pham, Phys. Lett. A{\bf 250},
25(1998), ibid, {\bf 250}, 29(1998)

\bibitem{ref9}  C. M. Bender, K. A. Milton, V. M. Savage, Phys. Rev. D{\bf %
62}, 085001(2000),\newline C. M. Bender, S. Boettcher, H. F.
Jones, P. N. Meisinger, J. Math. Phys. {\bf 42},
1960(2001),\newline C. M. Bender, S. Boettcher, H. F. Jones, P. N.
Meisinger, M. \c{S}im\c{s}ek, Phys. Lett. A{\bf 291},
197(2001),\newline C. M. Bender, Czech. J. Phys. {\bf 54},
13(2004)

\bibitem{ref10}  B. Bagchi, C. Quesne, Phys. Lett. A{\bf 273}, 285(2000) 285

\bibitem{ref11}  B. Bagchi, C. Quesne, Phys. Lett. A{\bf 300}, 18(2002)

\bibitem{ref12}  G. L\'{e}vai, F. Cannata, A. Ventura, J.Phys. A{\bf 34},
839 (2001)

\bibitem{ref13}  G. L\'{e}vai, F. Cannata, A. Ventura, J. Phys. A{\bf 35},
5041 (2002)

\bibitem{ref14}  F. Cooper, A. Khare, U. Sukhatme, Phys. Rep. {\bf 251},
267(1995), hep-th/0209062

\bibitem{ref15}  F. Cannata, G. Junker, J. Trost, Phys. Lett. A{\bf 246}
(1998) 219

\bibitem{ref16}  B. Bagchi, S. Mallik, C. Quesne, Mod. Phys. Lett. A{\bf 17}%
, 1651(2002)

\bibitem{ref17}  B. Bagchi, R. Roychoudhury, J. Phys. A{\bf 33} , L1(2000)

\bibitem{ref18}  A. A. Adrianov, et al., Int. J. Mod. Phys. A{\bf 14},
2675(1999)

\bibitem{ref19}  S. Vincenzo, V. Alonso, Phys. Lett. A{\bf 298} , 98(2002)

\bibitem{ref20}  J. S. Petrovi\'{c}, V. Milanovi\'{c}, Z. Ikoni\'{c}, Phys.
Lett. A{\bf 300}, 595(2002)

\bibitem{ref21}  F. M. Fernandez, R. Guardiola, J. Ros, M. Znojil, J. Phys.
A{\bf 31}, 10105(1998),\newline G. A. Merzinescu, J. Phys. A{\bf
33}, 4911(2000),\newline O. Mustafa, M. Znojil, J. Phys. A{\bf
35}, 8929(2002),\newline M. Znojil, F. Gemperle, O. Mustafa, J.
Phys. A{\bf 35}, 5781(2002),\newline M. Znojil, J. Phys. A{\bf
35}, 2341(2002),\newline M. Znojil, Phys. Lett. A{\bf 271},
327(2000)

\bibitem{ref22}  V. Milanovi\'{c}, Z. Ikoni\'{c}, Phys. Lett. A{\bf 293},
29 (2002)

\bibitem{ref23}  C. S. Jia, X. L. Zeng, L. T. Sun, Phys. Lett. A{\bf 294},
185 (2002),\newline M. Znojil, J. Phys. A{\bf 36},
7639(2003),\newline M. Znojil, J. Phys. A{\bf 36},
7825(2003),\newline M. Znojil, G. L\'{e}vai, Mod. Phys. Lett.
A{\bf 16}, 2273(2001),\newline M. Znojil, Phys. Lett. A{\bf 285},
7(2001), hep-th/0404213

\bibitem{ref24}
H. Ta\c{s}eli, J. Phys. A {\bf 31}, 779 (1998),
\newline
M. Znojil, Phys. Lett. A{\bf 264} (1999) 108,\newline
\"{O}. Ye\c{s}ilta\c{s}, M. \c{S}im\c{s}ek, R. Sever, C. Tezcan,
Phys. Scripta {\bf T67}, 472(2003),\newline D. T. Barclay, R.
Dutt, A. Gangopadhyaya, A. Khare, A. Pagnamenta, C. S. Jia, Y.
Sun, Y. Li, Phys. Lett. A{\bf 305}, 231(2002),\newline G.
L\'{e}vai, M. Znojil, J. Phys. A{\bf 35}, 8793(2002)

\bibitem{ref25}  C. M. Bender, S. Boettcher, J. Phys. A{\bf 31}, L273(1998),%
\newline
C. M. Bender, S. Boettcher, H. F. Jones, V. M. Savage, J. Phys.
A{\bf 32}, 6771(1999),\newline M. Znojil, J. Phys. A{\bf 33},
4203(2000),\newline A. Sinha, G. L\'{e}vai, P. Roy, Phys. Lett.
A{\bf 322}, 78(2004)

\bibitem{ref26}  C. M. Bender, G. V. Dunne, P. N. Meisinger, M. \c{S}im\c{s}%
ek, Phys. Lett. A{\bf 281}, 311(2001)

\bibitem{ref27}  A. Khare, U. Sukhatme, Phys. Lett. A{\bf 324}, 406(2004)

\bibitem{ref28}  G. L\'{e}vai, M. Znojil, J. Phys. A {\bf 33}, 7165(2000)

\bibitem{ref29}  Parthasarathi, D. Parashar, R. S. Kaushal, J. Phys. A{\bf %
37}, 781(2004)

\bibitem{ref30}  B. G\"{o}n\"{u}l, O. \"{O}zer, Y. Can\c{c}elik, M. Ko\c{c}%
ak, Phys. Lett. A{\bf 275}, 238(2000)

\bibitem{ref31}  E. Schr\"{o}dinger, Proc. R. Irish Acad. A{\bf 46},
9(1940); ibid, {\bf 46}, 183(1940); ibid, {\bf 47}, 53(1941)

\bibitem{ref32}  L. Infeld, T. E. Hull, Mod. Phys. {\bf 23}, 21(1951)

\bibitem{ref33}  M. Ioffe, A. I. Neelov, J. Phys. A{\bf 36}, 2493(2003)

\bibitem{ref34}  V. M. Tkachuk, T. V. Fityo, J. Phys. A{\bf 34}, 8673(2001)

\bibitem{ref35}  E. D. Filho, R. M. Ricotta, Phys. Lett. A{\bf 269},
269(2000)
\bibitem{ref36}  C. S. Lam, Y. P. Varshni, Phys. Rev. A {\bf 4}
(1971) 1875
\bibitem{ref37}  S. Fl\"{u}gge, Practical Quantum Mechanics, Springer,
Berlin (1974)
\end{thebibliography}
\end{document}